# THE DEFICIT OF DISTANT GALAXY CLUSTERS IN THE RIXOS X-RAY SURVEY

Francisco J Castander[1], Richard G Bower[2], Richard S Ellis[1],

Alfonso Aragón-Salamanca[1], Keith O Mason[3], Günther Hasinger[4,5],

Richard G McMahon[1], Francisco J Carrera[3], Jon P D Mittaz[3],

Ismael Pérez-Fournon[6] and Harry J Lehto[7]

[1] Institute of Astronomy, Madingley Road, Cambridge CB3 0HA, UK.
[2] Royal Observatory of Edinburgh, Blackford Hill, Edinburgh EH6 3HJ, UK.
[3] Mullard Space Science Laboratory, Holmbury St Mary, Dorking, Surrey RH5 6NT, UK.
[4] Max Planck Institut für Extraterrestrische Physik, D-8046 Garching, Germany.
[5] Astrophysikalisches Institut Postdam, An der Sternwarte 16, 14482 Potsdam, Germany.
[6] Instituto de Astrofísica de Canarias, 38200 La Laguna, Tenerife, Spain.
[7] Turku University Observatory, Vaisalantie 20, 21500 Tuorla, Piikkio, Finland.

Clusters of galaxies are the largest gravitationally bound systems and therefore provide an important way of studying the formation and evolution of the large scale structure of the Universe. Cluster evolution can be inferred from observations of the X-ray emission of the gas in distant clusters, but interpreting these data is not straightforward. In a simplified view, clusters grow from perturbations in the matter distribution: their intracluster gas is compressed and shock-heated by the gravitational collapse[1]. The resulting X-ray emission is determined by the hydrostatic equilibrium of the gas in the changing gravitational potential. However, if processes such as radiative cooling or pre-collapse heating of the gas are important, then the X-ray evolution will be strongly influenced by the thermal history of the gas. Here we present the first results from a faint flux-limited sample of X-ray selected clusters compiled as part of the ROSAT International X-ray and Optical Survey (RIXOS). Very few distant clusters have been identified. Most importantly, their redshift distribution appears to be inconsistent with simple models based on the evolution of the gravitational potential. Our results suggest that radiative cooling or non-gravitational heating of the intracluster gas must play an important role in the evolution of clusters.

Clusters of galaxies are visible as concentrations of physically-associated galaxies and as regions of extended X-ray emission. Optical surveys for distant clusters are, however, hampered by the likely superposition of physically separate systems along the line of sight. This makes it difficult to use such studies to reliably quantify cluster evolution. In contrast, X-ray studies are unaffected by such difficulties and thus offer a direct measure of the evolution. However, the flux limited surveys to date have only been able to detect the most luminous examples at cosmologically significant distances. Two groups[2,3] have reported a decrease in the amplitude of the X-ray luminosity function (XLF) with redshift. Edge et al[2], studying an ensemble sample from several surveys, found a deficit in the volume density of luminous clusters in the range $z \sim 0.1$–$0.2$ when compared with predictions based on the local XLF. The Einstein Extended Medium Sensitive Survey (EMSS)[3,4] showed a similar trend at a fainter limit. If we accept that the power spectrum of cosmological density fluctuations is similar to that of the Cold Dark Matter model[5], these results are in clear contradiction with the simplest models of evolution, where the intracluster gas evolves with the cluster potential in a self-similar fashion. In such models the typical cluster masses decrease with redshift while their densities increase. The X-ray luminosity is more strongly related to the density (the balance being set by the choice of power spectrum) and therefore the net prediction is a higher number of clusters of a given luminosity in the past or, conversely, an increase in the amplitude of the XLF with redshift. Intriguingly, the EMSS data for the highest redshift bin sampled ($\bar{z}$=0.33) show a steeper slope in the XLF. If this trend were maintained to lower luminosities, yet fainter surveys would reveal a large increase in the number of sub-luminous clusters at intermediate redshifts over expectations based on the local XLF. The fainter flux limit achievable with ROSAT[6] can thus test the *form* of cluster X-ray evolution as well as to extend previous tests to higher redshift.

The sample presented here was derived from the ROSAT International X-ray Optical Survey (RIXOS), a project aimed at producing a catalogue of optical identifications for $\simeq$400 X-ray sources found in 81 northern ROSAT fields observed with the Position Sensitive Proportional Counter (PSPC). Fields at high Galactic latitude ($b > +28°$) were selected with exposure times $> 8$ Ksec achieving a limiting flux optimised for wide-area optical follow-up. To avoid degradation of the point spread function at large radii, and interference from the PSPC window support structure, only sources within $17'$ of the field centre were chosen. In total, 385 X-ray sources were catalogued in 81 fields over 20.4 deg$^2$ to a limiting flux of $f_X \geq 3.0 \times 10^{-14}$ erg s$^{-1}$cm$^{-2}$ in the 0.5–2.0 keV energy band (Carrera et al, in prep).

Optical imaging and spectroscopic follow-up was carried out at La Palma using time allocated under the international programme supported by the Comité Científico Internacional. The results presented here refer to a subset of 59 randomly-selected fields for which complete identifications were secured for 264 sources. In considering completeness for the analysis below, allowance has been made for a few bright ($2^m$–$14^m$) stars very close to the X-ray position whose low resolution spectra (9Å) are insufficient to reveal evidence of chromospheric activity. A further 14 sources (5%) have no convincing spectroscopic identification; however in all but three cases, there is only a single isolated candidate (presumably a BL Lac) within the X-ray error box. For the remaining 3 sources, only one could conceivably be a cluster; a deep image in good conditions reveals a number of very faint sources unfortunately beyond reach of 4-m spectrographs.

Fig. 1a shows the redshift distribution of the 13 identified RIXOS clusters with redshift greater than 0.2. As discussed above, there is at most one distant candidate which we tentatively place at $z=1$ on the basis of the apparent $R$ magnitudes of the brightest galaxies. Regardless of this uncertainty in the high $z$ tail, we demonstrate that pushing to a fainter flux limit has neither revealed an abundant population of distant luminous clusters *nor* numerous sub-luminous clusters at intermediate redshift.

Fig. 1a also shows predicted redshift distributions for the RIXOS selection criteria calculated by integrating an appropriate XLF over luminosity and redshift in 3 physically-interesting cases. Each prediction incorporates a maximum likelihood fitting of the cluster emission to yield the detected flux. The model profile is determined by convolving a standard $\beta = \frac{2}{3}$ profile[7] (with a core radius of $r_0 = 250\,\mathrm{Kpc}$) with a Gaussian approximation of the PSPC point spread function. Calculations are corrected to the 0.5–2.0 keV band, assume $H_0 = 50\,\mathrm{km\,s^{-1}\,Mpc^{-1}}$ and $q_0 = 0.5$. Table 1 compares, for the three cases, the relative likelihoods of accounting for the data obtained in this survey. A unique feature of the RIXOS survey, however, arising from its faint flux limit, is its ability to distinguish between different evolutionary models which are equally consistent with the earlier EMSS data observed to a brighter flux limit. To illustrate this point, we also compare our evolutionary predictions with published data for the EMSS survey[4,8] in Fig. 1b. Table 2 compares the areas covered and the flux limits achieved in both surveys.

Our first prediction is that expected for a non-evolving cluster population based upon Edge et al's local XLF[2]. Although this model is inconsistent with the earlier X-ray results[2,3], it is still an important null hypothesis given that optical surveys[9] have failed to detect any strong evolution in the volume density of rich clusters to $z \sim 0.5$. However, in this case, given our faint limiting flux, we ought to readily detect X-ray clusters to very high redshifts. Even including the tentative candidate at $z \sim 1$, the Poisson probability of finding a single cluster beyond $z = 0.6$ when 7.9 are expected in the no-evolution case is 0.3%. Furthermore, the model is in qualitative disagreement with the steep decline in the number of clusters observed within $0.2 < z < 0.7$. The RIXOS survey therefore confirms the marked decline in the volume density of luminous clusters with redshift seen in the earlier surveys.

The next model we consider assumes that the slope of the initial spectrum of density fluctuations is shallow ($n \approx -2$). In this case, the initial density perturbations that lead to the formation of clusters are small and the gravitational collapse of the most massive clusters has occurred only recently. At high luminosities, this model produces a decrease in the number of high redshift clusters, but this is counter-balanced by a rapid increase (with redshift) in the numbers of sub-luminous clusters. To generate the model prediction we apply the self-similar scaling laws described by Kaiser[10]. These laws encapsulate the physical principles governing the evolution of cluster gravitational potentials. Thus, unless the intracluster gas responds to additional physical constraints, we can use self-similarity to accurately calculate the high-redshift XLF by rescaling and renormalising the present-day equivalent of Edge et al[2]. This self-similar scenario does not naturally reproduce the observed X-ray luminosity-temperature relation[11]; however, this model (with the flat spectral index) can account for the evolutionary trends seen in the EMSS survey. Significantly, the fainter RIXOS survey does not detect as many intermediate redshift clusters as expected in this model. Specifically, 32.9 clusters would be expected within $0.2 < z < 0.7$ compared with only 13 observed, a probability of 0.01%.

Our final model examines the situation if the self-similarity of Kaiser's scaling laws XLF is broken. Physical processes such as gas cooling, gas stripping, feedback from galaxy formation or winds could each break this scaling. In an open Universe, it is possible that the changing rate at which gas falls into the cluster potential could produce a similar effect. Within the variety of possible evolutionary scenarios that might be proposed, we choose one which is intended to reproduce qualitatively the 'preheated' intracluster medium model proposed by Kaiser[10]. In such a model the intracluster gas is heated at some early epoch acquiring a high characteristic entropy. This might arise, for example, via multiple supernova explosions occurring at the epoch of galaxy formation which expel energy into the primordial intracluster medium. Thereafter the central gas, which produces most of the emission, remains adiabatic through the collapse process. The resulting X-ray luminosity-temperature relation has a slope consistent with its present-day value[11] which does not depend on redshift. We consider the spectral index of the density fluctuations as a free parameter which is adjusted to give the best fit to the data. With a spectral index of $n = -1.5$, a good fit to the RIXOS results can be obtained (Table 1). A likelihood ratio test[12] based on the RIXOS data with $0.2 < z < 0.7$ shows an almost twenty-fold improvement for the 'fixed $L_X$–$T_X$' model over the 'self-similar' case (formally corresponding to greater than 99.99% confidence). Likewise, the ratio compared to the no evolution case is 8.7 (99.7% confidence). We note a further important point in this comparison: a different spectral parameter is required to fit the RIXOS and EMSS data sets (see Table 1). This most probably reflects the greater fragility of low luminosity clusters to the effects of preheating — a subtlety that has not been incorporated into our simple model.

Notwithstanding the small sample obtained in the RIXOS survey compared to the extensive EMSS survey, by probing to fainter flux limits our data demonstrate two important results. Firstly, there remains a substantial deficit of high redshift clusters compared to no evolution expectations confirming and strengthening conclusions based on earlier studies. Secondly, at intermediate redshifts we have penetrated sufficiently far down the XLF to distinguish between two evolutionary scenarios which were both consistent with the previous data. We do not observe the expected increase in the number density of sub-luminous clusters predicted in the case where the XLF evolves according to a simple self-similar scaling law. It appears that an additional hydrodynamical ingredient is required to break this scaling. By defining the form of the evolution of the X-ray luminosity function, our data open the way for a theoretical picture that properly accounts for the thermal history of the intracluster gas. One appealing possibility is that the intra-cluster gas was heated at an early epoch (perhaps as the result of the galaxy formation process) and has only recently become sufficiently cool to be trapped by the cluster gravitational potential.

**ACKNOWLEDGEMENTS.** The RIXOS project has been made possible by the the award of International Time on the La Palma telescopes by the Comité Científico Internacional. We thank numerous individuals who have contributed and acknowledge stimulating discussions with Carlos Frenk. F J Castander acknowledges financial support from the Instituto de Astrofísica de Canarias and hospitality from the LAEFF.

# TABLE 1

# Relative Likelihood Values ($0.2 < z < 0.7$)

| Model | RIXOS Likelihood | EMSS Likelihood |
|---|---|---|
| Self-Similar ($n \simeq -2$) | 19.3 | 0.0 |
| No Evolution | 8.7 | 9.2 |
| Fixed $L_X$–$T_X$ | 0.0 ($n = -1.5$) | 0.4 ($n = -0.8$) |

# TABLE 2

## RIXOS and EMSS flux limits and area coverages

| Survey | Flux limit $10^{-14}$ erg cm$^{-2}$ s$^{-1}$ | area coverage deg$^2$ |
|--------|-----------|---------------|
| RIXOS  | 3.0       | 14.9          |
| EMSS   | 3.6       | 0.09          |
| EMSS   | 6.2       | 6.4           |
| EMSS   | 7.4       | 15.1          |

Comparison of the RIXOS and EMSS surveys. The entire area covered by RIXOS was surveyed to the same limiting flux in the ROSAT 0.5-2.0 KeV band. However, the area covered by the EMSS survey was searched at different depths in the EINSTEIN 0.3-3.5 KeV band. For this comparison the EMSS flux limits[13] have been converted to the ROSAT band. It is important to note that the relative areas surveyed depend on the cluster temperature and redshift assumed, and are also affected by different detection procedures which are similarly redshift dependent. This table has been calculated for a cluster of temperature $T_x = 3$ KeV at a redshift of $z = 0.3$. For the same area coverage, RIXOS is a factor 2.5 deeper than the EMSS in this particular case.

**Figure Caption**

**Figure 1:** Redshift distribution for complete samples of faint X-ray selected clusters from (a) the RIXOS and (b) the EMSS[4,8] surveys. The dotted datum in the RIXOS survey represents a candidate cluster with an assigned redshift of $z = 1.0$ (see text). Smooth curves show the predicted redshift distribuitons for several evolutionary scenarios: (i) a non-evolving cluster population (based on Edge et al's local XLF[2]); (ii) a self-similar model with $n = -1.95$ spectral index; and (iii) the best fitting preheated intra-cluster medium model (with spectral index parameter $n = -1.5$ and $n = -0.8$ in (a) and (b) respectively). Although both surveys have detected nearby clusters, only clusters with redshift $z > 0.2$ are presented in the figure. For clusters at lower redshifts, uncertainties in modelling the cluster profile and the XLF at very low luminosities ($L_x \leq 5 \times 10^{42}$ erg s$^{-1}$) make comparison with model predictions unreliable. Despite the smaller sample of the RIXOS survey (due its smaller total area coverage), RIXOS is able to detect lower luminosity clusters (cf., Table 2). In this way, it probes the evolution of the faint end of the XLF enabling the distinction between evolutionary models that were both compatible with the earlier EMSS survey.

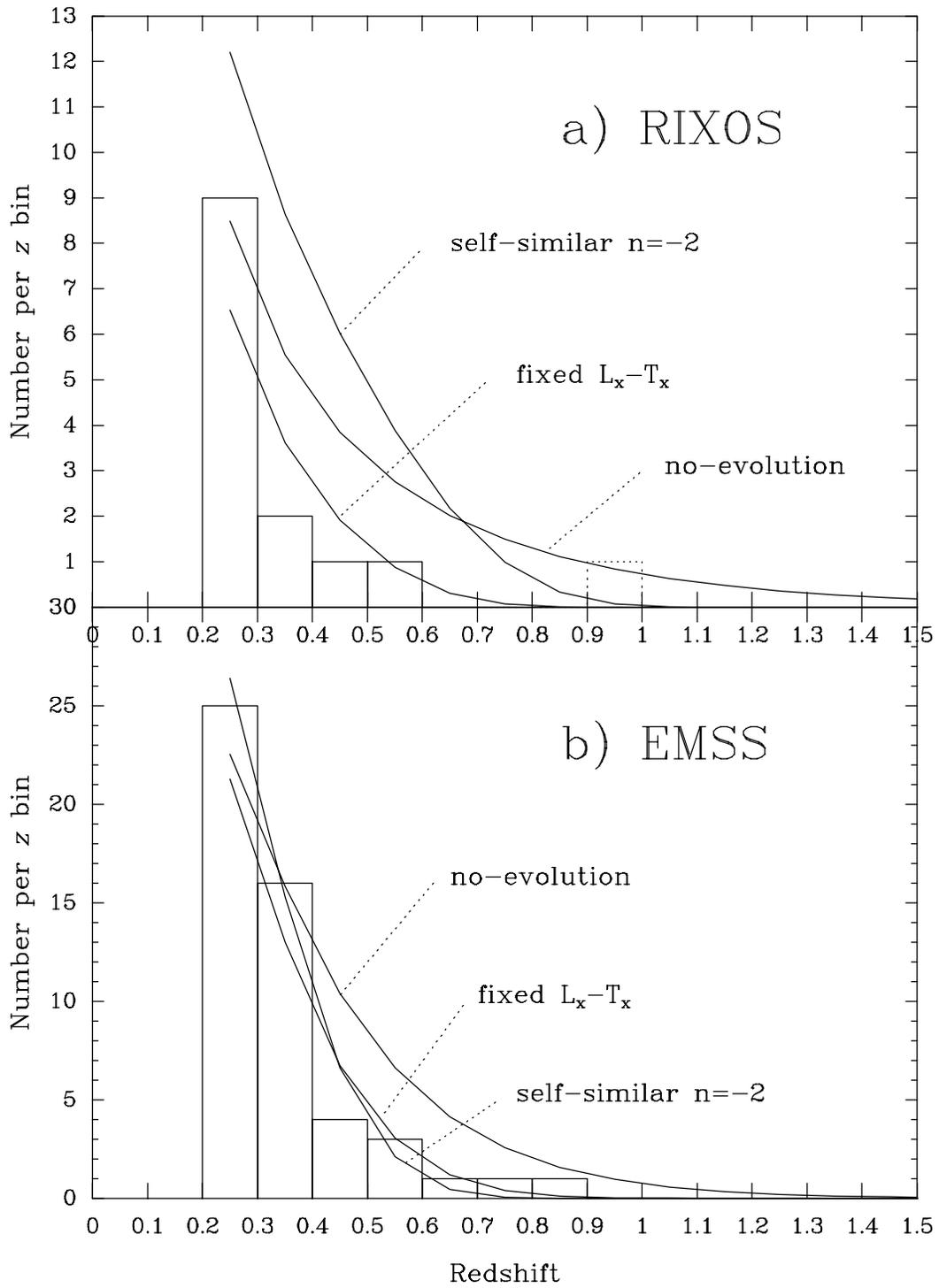

FIGURE 1